\documentclass[aps,pra,twocolumn,epsf,showpacs,superscriptaddress]{revtex4}
\usepackage{graphicx}
\usepackage{amsmath,amssymb,latexsym}
\usepackage{bm}

\begin{document}

\title{Action principle for cellular automata and the linearity of quantum mechanics} 

\author{Hans-Thomas Elze}
\affiliation{Dipartimento di Fisica ``Enrico Fermi'',  
        Largo Pontecorvo 3, I-56127 Pisa, Italia } 

\email{elze@df.unipi.it}

\begin{abstract} 
We introduce an action principle for a class of integer valued cellular automata   
and obtain   
Hamiltonian equations of motion. Employing sampling theory, these  discrete  deterministic equations are invertibly mapped on continuum equations for a set of bandwidth limited harmonic oscillators, which  
encode the Schr\"odinger equation. 
Thus, the linearity of quantum mechanics 
is related to the action principle of such cellular automata and 
its conservation laws to discrete ones. 
\end{abstract}
\maketitle 

\section{Introduction} 
The linearity of quantum mechanics (QM) is obvious in the Schr\"odinger equation 
and similarly in its functional form in quantum field theory (QFT). It is a fundamental aspect that does not depend on the object under study, if it is sufficiently isolated from 
anything else. Most importantly, by way of the superposition principle, linearity 
entails such ``quantum essentials'' as interference and entanglement. 
They are at the core of QM and of its applications alike, 
{\it e.g.} in advanced precision measurement and information technologies.    

Nevertheless, linearity of QM has been questioned from time to time and particular nonlinear modifications have been proposed. They have been subjected to experimental tests, putting bounds on their parameters when none of their   
predicted effects have been seen. Ample discussion and a stepwise proof that QM has to be linear have been provided by Jordan, based on the separability  
assumption {\it ``... that the dynamics we are considering can be independent
of something else in the universe, that the system we are considering can be described as
part of a larger system without interaction with the rest of the larger system.''} 
\cite{Jordan}. In Weinberg's articles, {\it e.g.}, a class of modifications 
and their relation to experimental signatures have been studied \cite{Weinberg}. In this case, theoretical objections have been raised, 
showing that the proposed nonlinearities would lead to superluminal 
signals or communication between branches of the wave function  \cite{Gisin, Polchinski}; 
since then, ``no signalling'' has become a versatile criterium confronting attempted modifications of QM \cite{me08}. 

The purpose of this article is to demonstrate a relation between QM and mechanics of a class of Hamiltonian cellular automata. In this way, the linearity of QM becomes an unavoidable  feature deriving from the action principle governing the discrete dynamics.

This is motivated by explorations of discrete deterministic mechanics by Lee \cite{Lee}, by the study of bandwidth limited fields and their possible role in 
discrete structures on and of spacetime by Kempf \cite{Kempf},  and by the representation of QM in 
terms of classical notions of observables, phase space, and Poisson bracket algebra  
by Heslot \cite{Heslot85,me12}. A combination of these ideas promises to be  
fruitful for our  understanding of interference, entanglement, and    
measurement in QM and for new approximation schemes in quantum theory.  

\section{Discrete Hamiltonian mechanics}
Discreteness arises in many contexts in physics 
or mathematics, besides quantization, {\it e.g.},  
in discrete maps facilitating numerical studies of complex systems, as regularized versions of quantum field theories on spacetime lattices, or 
describing intrinsically discrete processes. Finite difference equations 
are expected to play a role here, diminishing the preponderance of differential 
calculus.   

Lee and collaborators proposed to incorporate  fundamental discreteness into all of dynamics \cite{Lee} (and references therein), 
witnessing the difficulties in trying to find a 
manageable theory of ``quantum gravity'', let alone ``the'' unified theory.   
Deterministic discrete mechanics 
derives from the assumption that {\it time is a discrete dynamical variable}. This 
invokes a {\it fundamental length or time scale} (in natural units), $l$, and can be more aptly phrased that in a fixed $(d+1)$-dimensional spacetime volume $\Omega$ maximally $\Omega /l^{d+1}$ measurements can be performed or this number of events take place 
\cite{Lee}. 

Here we consider the introduction of {\it one} additional discreteness scale $l$ -- 
a time scale being related to a length scale by  the  velocity 
of light in vacuum -- in the spirit of widespread interest in {\it deformations} of 
Lorentz symmetry, in the form of ``doubly special relativity'' (DSR) \cite{Hossenfelder},  or in its explicit breaking, 
and in the nonlinear {\it deformation} of QM 
\cite{Weinberg,Gisin,Polchinski,me08} mentioned in Sect.\,I., etc. While such studies 
necessarily introduce additonal parameters, the aim is to probe the stability of 
existing theories, such as QM and the Standard Model, against such deformations, 
in order to find windows where new phenomena might show up that eventually 
lead  to a deeper theory with a smaller set of fundamental parameters 
\cite{DuffOkunVeneziano}. In the absence of such unifying theory, it is commonly  
expected that $l\equiv l_{Pl}$, i.e., that discreteness and Planck scale coincide 
\cite{Hossenfelder}. 
 
Various discrete models have been elaborated, which share desirable symmetries 
with the corresponding continuum theories  while presenting finite degrees of freedom. 
Different forms and (dis)advantages 
of a Lagrangian formulation  \cite{Lee} have been discussed, 
{\it e.g.}, in 
Refs.\,\cite{DInnocenzo87,HongBin05,Toffoli11}. Instead, we introduce an  action  
principle which leads to particularly transparent and symmetric Hamiltonian equations 
of motion. This induces a discrete phase space picture of the dynamics. 
 
Consider a classical cellular automaton (CA) with a denumerable set of degrees of freedom and represent its state by {\it integer valued} ``coordinates''   
$x_n^\alpha ,\tau_n$ and ``conjugated momenta'' $p_n^\alpha ,\pi_n$, where 
$\alpha\in {\mathbf N_0}$ denote different degrees of freedom and $n\in {\mathbf Z}$  
different states. 

The {$x_n$ and $p_n$ might be higher dimensional vectors, while $\tau_n$ and 
${\cal P}_n$ are assumed onedimensional. In separating the ``coordinate'' $\tau_n$ 
from the $x_n^\alpha$'s (and correspondingly $\pi_n$ from the $p_n^\alpha$'s), 
we follow Refs.\,\cite{Lee,DInnocenzo87,HongBin05}, in order to indicate that one of 
the degrees of freedom 
is special in that it represents the {\it dynamical time} variable here,  unlike   
the external parameter time of standard mechanics or QM. 
   
Finite differences, for all dynamical variables, are defined by: 
\begin{equation}\label{findiff}
\Delta f_n:=f_n-f_{n-1} 
\;\;. \end{equation} 
Furthermore, we define (using henceforth the summation convention for Greek indices, 
$r^\alpha s^\alpha\equiv\sum_\alpha r^\alpha s^\alpha$): 
\begin{eqnarray}\label{CAHamiltonian} 
{\cal A}_n&:=&\Delta \tau_n (H_n+H_{n-1})+a_n 
\;\;, \\ [1ex] \label{CAHamiltonian1}
H_n&:=&\frac{1}{2}S_{\alpha\beta}(p_n^\alpha p_n^\beta+x_n^\alpha x_n^\beta  )
+A_{\alpha\beta}p_n^\alpha x_n^\beta + R_n 
\;\;, \\ [1ex] \label{CAHamiltonian2}
a_n&:=&c_n\pi_n
\;\;, \end{eqnarray}  
where constants, $c_n$, and symmetric, $\hat S\equiv\{ S_{\alpha\beta}\}$,  and antisymmetric, $\hat A\equiv\{ A_{\alpha\beta}\}$, matrices are all integer valued;  
$R_n$ stands for higher than second powers in $x_n^\alpha$ or  $p_n^\alpha$.  
The choice of $a_n$  influences the behaviour of the variable $\tau_n$; 
we consider here only a very simple possibility \cite{time}. 
 
Defining the integer valued CA {\it action}:  
\begin{equation}\label{action} 
{\cal S}:=
\sum_n[(p_n^\alpha +p_{n-1}^\alpha )\Delta x_n^\alpha 
+(\pi_n+\pi_{n-1})\Delta\tau_n
-{\cal A}_n]  
\;\;, \end{equation} 
the evolution of the CA is determined by this: \\ 
{\it Postulate:} \hskip 0.1cm The CA follows the discrete updating rules 
(equations of motion)   
which  are determined by the {\it action principle} $\delta {\cal S}=0$,  
referring to arbitrary integer valued variations of all dynamical variables defined by: 
\begin{equation}\label{variation} 
\delta g(f_n):=[g(f_n+\delta f_n)-g(f_n-\delta f_n)]/2 
\;\; , \end{equation} 
where $f_n$ stands for one of the variables on which  polynomial $g$ may depend.\,$\Box$  
 
Several remarks are in order here. -- We observe that the variations  of 
constant, linear, or quadratic terms  yield results that are analogous to the continuum 
case. -- While infinitesimal variations do not conform with integer valuedness, 
there is no {\it a priori}  constraint on integer ones. However, for arbitrary  
$\delta f_n$, the {\it remainder of higher powers} in 
Eq.\,(\ref{CAHamiltonian1}), which enters the action, has to vanish for consistency, 
$R_n\equiv 0$. Otherwise the number of equations of motion 
generated by the action principle, generally, would exceed the number of 
variables \cite{initialconds}.     
 
Introducing the notation $\dot O_n:=O_{n+1}-O_{n-1}$,  
the following CA {\it equations of motion} are obtained:   
\begin{eqnarray}\label{xdotCA} 
\dot x_n^\alpha &=&\dot\tau_n(S_{\alpha\beta}p_n^\beta +A_{\alpha\beta}
x_n^\beta ) 
\;\;, \\ [1ex] \label{pdotCA} 
\dot p_n^\alpha &=&-\dot\tau_n(S_{\alpha\beta}x_n^\beta -A_{\alpha\beta}p_n^\beta ) 
\;\;, \\ [1ex] \label{taudotCA} 
\dot\tau_n&=&c_n 
\;\;, \\ [1ex] \label{pidotCA} 
\dot\pi_n&=&\dot H_n 
\;\;, \end{eqnarray} 
which are discrete analogues of Hamilton's equations, where all terms are defined in
terms of integers. The discrete {\it automaton time} $n$ is reflected by the    
finite difference equations here.  

Note that  the $\dot\tau_n$ 
present {\it background} parameters for the evolving $x,p$-variables, as  
a consequence of Eqs.\,(\ref{CAHamiltonian2}), (\ref{taudotCA}). 
Generally, $\dot\tau$ is a    
{\it lapse function} in Eqs.\,(\ref{xdotCA})--(\ref{pdotCA}).  

The Eqs.\,(\ref{xdotCA})--(\ref{pidotCA}) are {\it time reversal invariant}; 
the state $n+1$ can be calculated from knowledge of the 
earlier states $n$ and $n-1$ and the state $n-1$ from the later ones $n+1$ and $n$. 
 
Furthermore, there are conservation laws that are 
always respected by Eqs.\,(\ref{xdotCA})--(\ref{pdotCA}). -- Introducing the   
self-adjoint matrix $\hat H:=\hat S+i\hat A$,  
these equations yield: 
\begin{equation}\label{discrS} 
 \dot x_n^\alpha +i\dot p_n^\alpha =-i\dot\tau_n H_{\alpha\beta}
(x_n^\beta +ip_n^\beta ) 
\;\;, \end{equation} 
and its adjoint. Thus, we recover a {\it discrete analogue of Schr\"odinger's equation}, 
with $\psi_n^\alpha :=x_n^\alpha +ip_n^\alpha$ as the amplitude of the 
``$\alpha$-component''  
of ``state vector'' $|\psi\rangle$ at ``time'' $n$. Then, the 
Eqs.\,(\ref{xdotCA})--(\ref{pdotCA}) imply this:  \\ \noindent 
{\it Theorem A:} \hskip 0.1cm For any matrix $\hat G$ that commutes with $\hat H$, 
$[\hat G,\hat H]=0$, there 
is a {\it discrete conservation law}: 
\begin{equation}\label{Gconserv} 
 \psi_n^{\ast\alpha}G_{\alpha\beta}\dot\psi_n^\beta +
\dot\psi_n^{\ast\alpha}G_{\alpha\beta}\psi_n^\beta =0 
\;\;. \end{equation}  
For self-adjoint $\hat G$,  with complex integer elements, 
this relation concerns real integer quantities.\,$\Box$  \\  
{\it Corollary A:} \hskip 0.1cm For $\hat G:=\hat 1$, the Eq.\,(\ref{Gconserv}) implies 
a {\it conserved constraint} on the state variables:  
\begin{equation}\label{psiconserv} 
 \psi_n^{\ast\alpha}\dot\psi_n^\alpha +
\dot\psi_n^{\ast\alpha}\psi_n^\alpha =0 
\;\;. \end{equation}  
For $\hat G:=\hat H$, an {\it energy conservation} law follows.\,$\Box$  

Such matrices $\hat G$ generate    
{\it discrete unitary symmetry transformations}; admissible ones    
preserve complex  integer valuedness of the 
CA variables $\psi_n^\alpha$. 

Note that Eqs.\,(\ref{Gconserv}) and (\ref{psiconserv}) {\it cannot} be trivially 
``integrated'', since the {\it Leibniz rule} is modified. Recalling  
$\dot O_n:=O_{n+1}-O_{n-1}$, we have, for example, 
$O_{n+1}O'_{n+1}-O_{n-1}O'_{n-1}=\frac{1}{2}(\dot  O_n[O'_{n+1}+O'_{n-1}]
+[O_{n+1}+O_{n-1}]\dot O'_n)$, instead of the product rule of differentiation. 

Furthermore, we cannot  obtain a continuum limit simply by letting  
the discreteness scale $l\rightarrow 0$, 
as for example in   Refs.\,\cite{Lee,me13}.   
Integer valuedness here conflicts with continuous time 
and related derivatives. 

It is worth recalling the underlying assumption of discrete mechanics that the density 
of events and, thus, of information content of spacetime regions 
is cut off by the scale $l$ \cite{Lee,SorkinDICE02}.  
We may wonder whether the discreteness of a deterministic CA  
can be reconciled with any continuum description at all and, in particular, with QM?  

\section{Sampling theory}
We propose an answer here by exploring 
the possibility that physical fields, wave functions in particular, could be {\it simultaneously 
discrete and continuous}, represented by sufficiently smooth functions containing a finite 
density of degrees of freedom. This idea has recently been introduced by Kempf 
and has led to constructing a covariant ultraviolet cut-off suitable for theories 
including gravity  --  with motivation provided by ubiquitous appearance of 
generalized uncertainty relations \cite{Kempf}.  
However, neither {\it integer valued CA} nor the {\it structure of QM} have been 
addressed in this context. 

In his pioneering work, Shannon pointed out that information can have simultaneously
continuous and discrete character \cite{Shannon}. This has become a matter 
of routine application in signal processing, whenever conversion between analog and 
digital encoding is needed. Sampling theory 
demonstrates that any bandlimited signal can be perfectly reconstructed, 
provided discrete samples of it are taken at the rate of at least twice the band 
limit (Nyquist rate). For an extensive review, see \cite{Jerri}; see also  \cite{StrohmerTanner}, referring  to modern ramifications of the theory. 

For our present purposes, the {\it Sampling Theorem} in its 
simplest form suffices \cite{Kempf,Jerri}:  
Consider square integrable {\it bandlimited functions} $f$, {\it i.e.}, which can be 
represented as $f(t)=(2\pi )^{-1}\int_{-\omega_{max}}^{\omega_{max}}\mbox{d}\omega\; 
\mbox{e}^{-i\omega t}\tilde f(\omega )$, with bandwidth $\omega_ {max}$. Given 
the set of amplitudes $\{ f(t_n)\}$ for the set  $\{ t_n\}$ of equidistantly spaced times  
(spacing $\pi /\omega_{max}$), the function $f$ is obtained for all $t$ 
by: 
\begin{equation}\label{samplingtheorem} 
f(t)=\sum_n f(t_n)\frac{\sin [\omega_{max}(t-t_n)]}{\omega_{max}(t-t_n)} 
\;\;. \end{equation} 

Since the CA time is given by the integer $n$, the corresponding discrete {\it physical time}  
is obtained by  multiplying with the fundamental scale $l$,  $t_n\equiv nl$, and the 
bandwidth by $\omega_{max}=\pi /l$. 

Attempting to {\it map invertibly} Eqs.\,(\ref{xdotCA})--(\ref{pdotCA}) on reconstructed 
continuum equations, according to Eq.\,(\ref{samplingtheorem}), the nonlinearity 
on the right-hand sides is problematic: the product of two 
functions, with bandwidth $\omega_ {max}$ each, is not a function with the same 
bandwidth. Therefore, the mapping can only be consistent, if $\dot\tau_n$ is a 
constant.  

Let us recall Eq.\,(\ref{discrS}). Inserting $\psi_n^\alpha :=x_n^\alpha +ip_n^\alpha$ and      
applying the {\it Sampling Theorem}, this discrete time equation
is mapped to the  {\it continuous time equation}: 
\begin{equation}\label{modS}
\frac{\hat D_l-\hat D_{-l}}{2}\psi^\alpha (t)=\sinh (l\partial_t)\psi^\alpha (t)
=\frac{1}{i}H_{\alpha\beta}\psi^\beta (t) 
\;, \end{equation} 
where we employed the translation operator defined by $\hat D_Tf(t):=f(t+T)$ and 
set $\dot\tau_n\equiv\dot\tau =2$  \cite{taudotvar}.  

Thus, we obtain the {\it Schr\"odinger equation}, however, modified in important ways. 
(We use QM terminology freely, while paying attention to new effects arising here.)  
The wave function $\psi^\alpha$ has bandwidth $\omega_{max}$, due to 
reconstruction formula (\ref{samplingtheorem}). This corresponds to an 
{\it ultraviolet cut-off} of the energy $E$ of stationary states of the generic form $\psi_E(t):=\exp (-iEt)\tilde\psi$. Indeed, diagonalizing the self-adjoint  Hamiltonian,  
$\hat H\rightarrow\mbox{diag}(\epsilon_0,\epsilon_1,\dots )$, the 
Eq.\,(\ref{modS}) yields the eigenvalue equation:
\begin{equation}\label{eigenS} 
\sin (E_\alpha l)=\epsilon_\alpha  
\;\;, \end{equation} 
and a {\it modified dispersion relation}, 
$E_\alpha =l^{-1}\arcsin (\epsilon_\alpha )=
l^{-1}\epsilon_\alpha [1+\epsilon_\alpha^{\;2}/3!+\mbox{O}(\epsilon_\alpha^{\;4})]$ 
\cite{disprel}.   
The spectrum $\{ E_\alpha\}$ is cut off by 
the condition $|\epsilon_\alpha |\leq 1$, entailing   
$|E_\alpha |\leq \pi /2l=\omega_{max}/2$, {\it i.e.} half the bandlimit.  


The modified Schr\"odinger equation (\ref{modS}) incorporates higher-order 
time derivatives. These are negligible for low-energy wave functions, which vary little with 
respect to the cut-off scale, {\it i.e.}    
$|\partial^k\psi /\partial t^k|\ll l^{-k}=(\omega_{max}/\pi )^k$. 
  
Furthermore, the relation between Eq.\,(\ref{discrS}) and Eq.\,(\ref{modS}), together 
with the linearity of both equations, suggest that the correct continuous time 
conservation laws are obtained by the replacement: 
\begin{equation}\label{cconserv}   
\dot\psi_n:=\psi_{n+1}-\psi_{n-1}\;\;\longrightarrow\;\;\frac{1}{i}\sin (il\partial_t)\psi (t)
\;\;, \end{equation} 
{\it cf.} Eqs.\,(\ref{Gconserv}) and (\ref{psiconserv}), respectively. Indeed, by Eq.\,(\ref{modS}),  
the following holds: \\ 
{\it Theorem B:} \hskip 0.1cm For any matrix $\hat G$ that commutes with $\hat H$, 
there is a {\it continuous time conservation law}: 
\begin{equation}\label{cGconserv} 
 \psi ^{\ast\alpha}G_{\alpha\beta}\sin (il\partial_t)\psi^\beta +
[\sin (il\partial_t)\psi^{\ast\alpha}]G_{\alpha\beta}\psi^\beta =0 
\;\;, \end{equation}  
in particular,     
\begin{equation}\label{cpsiconserv} 
 \psi ^{\ast\alpha}\sin (il\partial_t)\psi^\alpha +
[\sin (il\partial_t)\psi^{\ast\alpha}]\psi^\alpha =0 
\;\;, \end{equation}  
which appropriately modifies the QM   
wave function {\it normalization}, referring to a basis denumerated by $\alpha$.\,$\Box$ 

The Eqs.\,(\ref{cGconserv})--(\ref{cpsiconserv})  allow us to remove the 
ultraviolet cut-off,  $l\rightarrow 0$, recovering QM results from the leading 
order terms. (If $l$ is a fundamental {\it constant}, this limit may be interesting for heuristic reasons alone.)   
-- For example, consider the real symmetric {\it two-time function}, 
\begin{equation}\label{C}
2C_{\hat G}(t_1,t_2):=\psi ^{\ast\alpha}(t_1)G_{\alpha\beta}\psi^\beta (t_2)\; +\;
\mbox{c.c.}
\;\;, \end{equation}
where $X+\mbox{c.c.}:=X+X^\ast$ and 
$\hat G$ is a self-adjoint matrix, with $[\hat G,\hat H]=0$. Inserting $C_{\hat G}$,        
{\it Theorem B} yields: \\ 
{\it Corollary B:} \hskip 0.1cm The two-time function $C_{\hat G}$ is invariant under 
discrete translations of this form: 
\begin{equation}\label{Ctransl} 
C_{\hat G}(t-l,t)=C_{\hat G}(t,t+l) 
\;\;, \end{equation} 
implying that it is fixed everywhere by giving  $C_{\hat G}(t,t+l)$ for all $t$ in an interval
$[t_0,t_0+l[$. \,$\Box$ \\ 
The wave function normalization,  
$\psi^{\ast\alpha}\psi^\alpha =1$,  
then arises here from the coincidence limit of a   
two-time function with the property $C_{\hat 1}(t,t+l)\equiv 1$, for all $t$: 
\begin{equation}\label{wfnorm} 
1=\lim_{l\rightarrow  0}C_{\hat 1}(t,t+l)=\psi^{\ast\alpha}(t)\psi^\alpha (t) 
\;\;, \end{equation} 
which is consistent with Eq.\,(\ref{cpsiconserv}) and  essential for the  probability interpretation in QM. An analogous {\it equal-time} constraint, in general,      
does not exist on the CA level of description. {\it E.g.},  
$\psi_n^{\ast\alpha}\psi_n^\alpha =x_n^\alpha x_n^\alpha +p_n^\alpha p_n^\alpha =1$, instead of Eq.\,(\ref{psiconserv}), is compatible only with rather trivial evolution, since 
all variables are integer valued.  
 
It is remarkable how properties of CA produce familiar QM results, even if modified 
by the finite scale $l$.  
Matrices that generate  QM conservation laws  
do so for the bandwidth limited continuum theory. Since the {\it same} vanishing commutator is responsible for CA conservation laws, Eqs.\,(\ref{Gconserv})--(\ref{psiconserv}), they strictly correspond to each other. Yet   
continuous QM symmetry transformations, generally, comprise a larger 
set than discrete CA ones 
respecting complex integer valuedness. 

\section{Discussion}
It will be interesting to find a similar one-to-one CA--QM map for 
relativistic QM and QFT. Since wave equations and functional 
Schr\"odinger equation are linear and have a Hamiltonian formulation, it should 
be possible to employ a generalized {\it Sampling Theorem} for fields.  
It has been shown how to covariantly regularize the d'Alembert operator by finite 
bandwidth of its spectrum \cite{Kempf}, which is a necessary ingredient. -- 
Conversely, given a {\it Hamiltonian} CA, we  may invoke the path integral 
for classical  systems  \cite{us}, with integration replaced by summation  
over integer valued variables, plus reconstruction formulae, in 
order to derive a relativistic bandwidth limited quantum (field) theory.   
-- Other constructions of CA for  
{\it relativistic models} have appeared, which either incorporate QM features from the  outset, {\it e.g.}, for the Dirac equation 
\cite{FeynmanIBBdArianoArrighi}, or derive them, {\it e.g.}, for bosonic QFT and 
a superstring model \cite{tHooft}; see also references there. 
These models have been all {\it noninteracting}. 
 
Lack of interactions there seems dictated by additional 
restrictions, such as locality, due to placing a CA, say at Planck scale, 
into physical spacetime as experienced at scales   
where QM is tested. Remarkably, arbitrary QM $N$-level systems 
can be described by $2N-1$ nonrelativistic coupled 
oscillators in one fictitious space dimension  \cite{Skinner13}. Are these hints  
that fundamental CA exist in an abstract space and that 
QM and spacetime emerge from there? 
 
The nonrelativistic CA considered in this article {\it do} incorporate {\it interactions} 
through  matrix elements $H_{\alpha\beta}$.  
Their $x^\alpha,p^\alpha$- variables can be embedded 
into twodimensional phase space, similarly as in Ref.\,\cite{Skinner13}. Yet  
other interpretations are possible, such as $\alpha$ labelling sites of a $d$-dimensional 
lattice or, generally, elements of a Hilbert space in   
the QM description following {\it sampling theory}. This freedom is due to 
the nonrelativistic formalism without reference to gravitation or   
dynamical spacetime.  

Another known approach allowing for   
interactions is a statistical theory of certain matrix models, which shows    
QM behaviour to emerge from a Gibbs distribution \cite{Adler}. 
Similarly as in Refs.\,\cite{tHooft}, however, this assumes a particular dynamics  
and it remains to 
be seen whether gauge theories as, for example,  in the Standard Model can 
be covered. Our approach here, in distinction, does not make assumptions 
about specific interactions or forces but explores a mapping between structural 
features of Hamiltonian CA and of QM. It will be challenging to identify the  
principles that govern a physically relevant Hamiltonian $\hat H$ within the 
``ontology'' of CA.    

It is worth while to also draw attention to the essential feature of {\it entanglement} 
in QM, 
as well as to the often discussed apparent nonlocality, and how this is reflected on 
the CA level in our approach. 

Considering the apparent nonlocality, we may refer here to recent work demonstrating  
that QM is a local theory in a well-defined sense, while much of ongoing 
debates must be attributed to terms which are imprecisely defined or 
used with varying connotations \cite{Englert}.  Concerning this, 
however, our mapping between CA and QM does {\it not} change QM, except by 
introducing the fundamental parameter $l$ in correction terms to Schr\"odinger 
equation and corresponding conservation laws. These modifications are negligible 
in the realm where QM has been tested, if $l$ belongs to the Planck scale. Therefore, 
arguments brought forth in the 
discussion of locality in QM, especially in Ref.\, \cite{Englert}, apply here to the same 
extent. 

Entanglement is present in our theory as in QM, since it arises as a consequence of 
its {\it linearity} embodied in the {\it superposition 
principle}, which has been main topic of this article. More specifically, there can be 
entangled states, when the relevant Hilbert space is a tensor product of subspaces. 
Which, in the simplest case, allows to have 
superposition ``Bell'' states, {\it e.g.}  
$|\psi\rangle =|\alpha\rangle\otimes |\beta\rangle \pm 
|\beta\rangle\otimes |\alpha\rangle$, where the first factor of the tensor products  
belongs to a subspace ``A'' and the second to a subspace ``B''. This type of structure, 
or its generalizations, can be built in our linear theory as well, and even on 
the CA level where the theory is again linear. We have not explicitly 
discussed tensorized spaces, when introducing wave function components 
$\psi^\alpha$ referring to a Hilbert space with denumerable basis, nor 
when introducing canonically conjugated $x^\alpha ,p^\alpha$-variables 
for CA. Which is motivated by the fact  
that the tensorized structure can always be embedded in a sufficiently 
large Hilbert space, such that it is {\it de facto} absent, but is reflected in 
a corresponding change of the algebra of observables. This has been   
elaborated in detail recently, because of the relevance for quantum information protocols. 
A concise exposition for the case of overall pure states is in 
Ref.\,\cite{Harshman}.  We conclude that our theory does not produce 
deviations from QM that would affect entanglement, which remains a manifestation 
of the linearity on both levels, CA as well as QM.  

It will be most interesting to reconsider questions of entanglement and locality, 
when a relativistic generalization of the present theory becomes available.  

Observables, measurements, and Born rule can be discussed in bandwidth limited 
theory with help of Heslot's work \cite{Heslot85} and implications 
for CA be considered elsewhere.  

For completeness, we point out the differences between the presently introduced 
Hamiltonian CA and so-called quantum cellular or {\it quantum lattice-gas automata} 
(QLGA). -- The QLGA have recently found much attention, since 
they are, by construction, discretizations of the Schr\"odinger equation 
\cite{Meyer97,Boghosian98}, cf. also \cite{FeynmanIBBdArianoArrighi}. 
Thus, they are of  great interest for potential applications of 
quantum computation, if it can be realized in practice. They are constructed 
in configuration space specifically to reproduce in the continuum limit a quadratic kinetic 
energy term in the Schr\"odinger equation. This involves judicious choice of 
transformation matrices, i.e. implicitly of dimensionless parameters \cite{Boghosian98}. 
In the absence of a physical guiding principle, the Hilbert space structure of QM state space (with complex wave functions) and linearity and unitarity of the evolution are 
to be incorporated  
{\it ab initio}. -- In these respects, our approach differs remarkably: It is 
based on {integer valued} dynamical variables and an underlying {action principle}. 
This  {\it implies} linearity and unitarity together with all conservation laws, 
which we have obtained explicitly for Hamiltonian CA, in the discrete and 
the continuous time description. 
Unlike the case of QLGA and in accordance with the discussion in Sect.\,II. of  
the role of the discreteness scale $l$, the Hamiltonian CA here 
provide a {\it discrete deformation} of QM that reduces to it for $l\rightarrow 0$. 

Our results also suggest to simulate complex QM systems by mapping 
on computer friendly {\it integer valued}  
Hamiltonian CA. The CA updates are {\it error free}. Introducing a bandwidth, 
rescaling the modified Schr\"odinger equation followed by mapping Hamiltonian matrix 
elements approximately on complex integer ones, and finite time effects 
produce errors to be explored, {\it cf.} last of Refs.\,\cite{FeynmanIBBdArianoArrighi}. 

We remark that Planck's constant  $\hbar$ does not interfere with such a map and 
remains independent here of the discreteness scale $l$. This can be illustrated  as 
follows. We write the Schr\"odinger equation in this form, $i\hbar\partial_{t'}\psi =
\epsilon_{phys}\hat h\psi$, where by $\epsilon_{phys}$ we factor out the 
physical energy scale of the problem at hand, such that the 
dimensionless Hamiltonian $\hat h$ is given by numbers that are (loosely speaking) 
``of O(1)''. Rescaling the time variable $t'/M'=:t$, with $M'\gg 1$, we obtain: 
$i\hbar\partial_t\psi =\epsilon_{phys}M'\hat h\psi
=(\hbar\omega_{max}/\pi )M\hat h\psi$, where we introduced the bandwidth 
limit, $\hbar\omega_{max}/\pi :=\epsilon_{phys}M'/M$, with $M'\gg M\gg 1$. 
At this point, units can be chosen such that $\hbar =1$, as usual. Furthermore, 
we introduce a complex integer valued Hamiltonian, $\hat H:=M\hat h$, 
as an approximation on the right-hand side of the Schr\"odinger equation, 
which may introduce errors for its matrix elements   
(loosely speaking) ``of O(1/$M$)''.  This presents the starting point for an 
analysis invoking sampling theory, in order to map the dynamics on a cellular 
automaton.   \\ 

\section{Conclusion}
In conclusion, a map between cellular automata (CA) and quantum mechanics (QM) 
has been constructed by a synthesis of 
elements from {\it discrete mechanics} \cite{Lee}, {\it sampling theory} \cite{Kempf}, 
and {\it Hamiltonian formulation}  of QM \cite{Heslot85}. QM can originate in  
integer valued CA incorporating a fundamental scale. The postulated action 
principle refers to (the only available) arbitrary integer variations of dynamical variables, 
which enforces the {\it linearity} of the theory. The {\it separability 
assumption} mentioned in Sect.\,I. before, which underlies an intrinsic derivation 
of the linearity of QM \cite{Jordan}, 
can be substituted by another statement in the present context: 
``... the dynamics we are considering can be independent of something else in the 
universe ...'', if and only if the relevant CA 
{\it action is stationary under arbitrary integer variations}.   
This may open another view of linearity and the superposition principle in 
quantum mechanics. 

\section*{Acknowledgements}
It is a pleasure to thank N. Buric and T. E. Skinner for critical reading and suggestions 
concerning a preliminary version of this paper, A. Kempf for feedback, and  an  
anonymous referee for careful remarks .   



\begin{thebibliography}{99}
\bibitem{time}  Additional `potential' terms can induce nontrivial behaviour 
of $\tau_n$; this has been considered in \cite{me13}.
\end{thebibliography}
\end{document}